\documentclass{PoS}

\usepackage{latexsym}
\usepackage{amsmath}
\usepackage{epsfig,graphics}

\PoS{PoS(LAT2005)256}

\title{Large N }

\ShortTitle{Large N }

\author{\speaker{Michael Teper}\\

        University of Oxford\\

        E-mail: \email{m.teper1@physics.oxford.ac.uk}}

\abstract{Some mysterious features of the strong interactions become
easily understood if our usual QCD with $N=3$ is `close to' 
SU($\infty$) and if the latter theory is confining. 
$N=\infty$ theories are theoretically simpler; in particular
there has been much progress in constructing weak-coupling duals 
in string theory. In this poster I will describe some of the
things that recent lattice calculations tell us about 
the large-$N$ limit of SU($N$) gauge theories in 3+1 dimensions. The 
focus is on confinement, how close SU($\infty$) is to SU(3), new stable 
strings at larger $N$, the Pomeron, deconfinement, topology, 't Hooft
string tensions. I also allude to other topics, such as the high-$T$
pressure deficit, chiral physics and the phases of the theory.}

\FullConference{XXIIIrd International Symposium on Lattice Field Theory\\

                 25-30 July 2005\\

                 Trinity College, Dublin, Ireland}

\begin{document}

\section{Some lattice results}

These calculations mostly proceed by looking at SU(2), SU(3), SU(4), SU(5),
.... and seeing if one can extrapolate to $N=\infty$ using the
expected  $O(1/N^2)$ correction.

\subsection{$N=3$ is close to $N=\infty$}

\begin	{figure}[ht]
\begin	{center}
\leavevmode
\setlength{\unitlength}{0.240900pt}
\ifx\plotpoint\undefined\newsavebox{\plotpoint}\fi
\sbox{\plotpoint}{\rule[-0.200pt]{0.400pt}{0.400pt}}%
\begin{picture}(1125,675)(0,0)
\font\gnuplot=cmr10 at 12pt
\gnuplot
\sbox{\plotpoint}{\rule[-0.200pt]{0.400pt}{0.400pt}}%
\put(300.0,150.0){\rule[-0.200pt]{4.818pt}{0.400pt}}
\put(275,150){\makebox(0,0)[r]{\ \ {$0$}}}
\put(1030.0,150.0){\rule[-0.200pt]{4.818pt}{0.400pt}}
\put(300.0,209.0){\rule[-0.200pt]{4.818pt}{0.400pt}}
\put(275,209){\makebox(0,0)[r]{\ \ {$1$}}}
\put(1030.0,209.0){\rule[-0.200pt]{4.818pt}{0.400pt}}
\put(300.0,269.0){\rule[-0.200pt]{4.818pt}{0.400pt}}
\put(275,269){\makebox(0,0)[r]{\ \ {$2$}}}
\put(1030.0,269.0){\rule[-0.200pt]{4.818pt}{0.400pt}}
\put(300.0,328.0){\rule[-0.200pt]{4.818pt}{0.400pt}}
\put(275,328){\makebox(0,0)[r]{\ \ {$3$}}}
\put(1030.0,328.0){\rule[-0.200pt]{4.818pt}{0.400pt}}
\put(300.0,388.0){\rule[-0.200pt]{4.818pt}{0.400pt}}
\put(275,388){\makebox(0,0)[r]{\ \ {$4$}}}
\put(1030.0,388.0){\rule[-0.200pt]{4.818pt}{0.400pt}}
\put(300.0,447.0){\rule[-0.200pt]{4.818pt}{0.400pt}}
\put(275,447){\makebox(0,0)[r]{\ \ {$5$}}}
\put(1030.0,447.0){\rule[-0.200pt]{4.818pt}{0.400pt}}
\put(300.0,506.0){\rule[-0.200pt]{4.818pt}{0.400pt}}
\put(275,506){\makebox(0,0)[r]{\ \ {$6$}}}
\put(1030.0,506.0){\rule[-0.200pt]{4.818pt}{0.400pt}}
\put(300.0,566.0){\rule[-0.200pt]{4.818pt}{0.400pt}}
\put(275,566){\makebox(0,0)[r]{\ \ {$7$}}}
\put(1030.0,566.0){\rule[-0.200pt]{4.818pt}{0.400pt}}
\put(300.0,625.0){\rule[-0.200pt]{4.818pt}{0.400pt}}
\put(275,625){\makebox(0,0)[r]{\ \ {$8$}}}
\put(1030.0,625.0){\rule[-0.200pt]{4.818pt}{0.400pt}}
\put(300.0,150.0){\rule[-0.200pt]{0.400pt}{4.818pt}}
\put(300,100){\makebox(0,0){\ {$0$}}}
\put(300.0,605.0){\rule[-0.200pt]{0.400pt}{4.818pt}}
\put(425.0,150.0){\rule[-0.200pt]{0.400pt}{4.818pt}}
\put(425,100){\makebox(0,0){\ {$0.05$}}}
\put(425.0,605.0){\rule[-0.200pt]{0.400pt}{4.818pt}}
\put(550.0,150.0){\rule[-0.200pt]{0.400pt}{4.818pt}}
\put(550,100){\makebox(0,0){\ {$0.1$}}}
\put(550.0,605.0){\rule[-0.200pt]{0.400pt}{4.818pt}}
\put(675.0,150.0){\rule[-0.200pt]{0.400pt}{4.818pt}}
\put(675,100){\makebox(0,0){\ {$0.15$}}}
\put(675.0,605.0){\rule[-0.200pt]{0.400pt}{4.818pt}}
\put(800.0,150.0){\rule[-0.200pt]{0.400pt}{4.818pt}}
\put(800,100){\makebox(0,0){\ {$0.2$}}}
\put(800.0,605.0){\rule[-0.200pt]{0.400pt}{4.818pt}}
\put(925.0,150.0){\rule[-0.200pt]{0.400pt}{4.818pt}}
\put(925,100){\makebox(0,0){\ {$0.25$}}}
\put(925.0,605.0){\rule[-0.200pt]{0.400pt}{4.818pt}}
\put(300.0,150.0){\rule[-0.200pt]{180.675pt}{0.400pt}}
\put(1050.0,150.0){\rule[-0.200pt]{0.400pt}{114.427pt}}
\put(300.0,625.0){\rule[-0.200pt]{180.675pt}{0.400pt}}
\put(150,537){\makebox(0,0){\Large{${{m_G}\over{\surd\sigma}}$}}}
\put(650,25){\makebox(0,0){\large{$1/N^2$}}}
\put(300.0,150.0){\rule[-0.200pt]{0.400pt}{114.427pt}}
\put(925.0,370.0){\rule[-0.200pt]{0.400pt}{2.168pt}}
\put(915.0,370.0){\rule[-0.200pt]{4.818pt}{0.400pt}}
\put(915.0,379.0){\rule[-0.200pt]{4.818pt}{0.400pt}}
\put(578.0,357.0){\rule[-0.200pt]{0.400pt}{1.927pt}}
\put(568.0,357.0){\rule[-0.200pt]{4.818pt}{0.400pt}}
\put(568.0,365.0){\rule[-0.200pt]{4.818pt}{0.400pt}}
\put(455.0,346.0){\rule[-0.200pt]{0.400pt}{1.686pt}}
\put(445.0,346.0){\rule[-0.200pt]{4.818pt}{0.400pt}}
\put(445.0,353.0){\rule[-0.200pt]{4.818pt}{0.400pt}}
\put(458.0,355.0){\rule[-0.200pt]{0.400pt}{3.132pt}}
\put(448.0,355.0){\rule[-0.200pt]{4.818pt}{0.400pt}}
\put(448.0,368.0){\rule[-0.200pt]{4.818pt}{0.400pt}}
\put(370.0,338.0){\rule[-0.200pt]{0.400pt}{2.409pt}}
\put(360.0,338.0){\rule[-0.200pt]{4.818pt}{0.400pt}}
\put(360.0,348.0){\rule[-0.200pt]{4.818pt}{0.400pt}}
\put(339.0,354.0){\rule[-0.200pt]{0.400pt}{3.373pt}}
\put(329.0,354.0){\rule[-0.200pt]{4.818pt}{0.400pt}}
\put(925,374){\circle*{12}}
\put(578,361){\circle*{12}}
\put(455,350){\circle*{12}}
\put(458,361){\circle*{12}}
\put(370,343){\circle*{12}}
\put(339,361){\circle*{12}}
\put(329.0,368.0){\rule[-0.200pt]{4.818pt}{0.400pt}}
\put(925.0,467.0){\rule[-0.200pt]{0.400pt}{3.132pt}}
\put(915.0,467.0){\rule[-0.200pt]{4.818pt}{0.400pt}}
\put(915.0,480.0){\rule[-0.200pt]{4.818pt}{0.400pt}}
\put(578.0,428.0){\rule[-0.200pt]{0.400pt}{2.650pt}}
\put(568.0,428.0){\rule[-0.200pt]{4.818pt}{0.400pt}}
\put(568.0,439.0){\rule[-0.200pt]{4.818pt}{0.400pt}}
\put(455.0,433.0){\rule[-0.200pt]{0.400pt}{3.132pt}}
\put(445.0,433.0){\rule[-0.200pt]{4.818pt}{0.400pt}}
\put(445.0,446.0){\rule[-0.200pt]{4.818pt}{0.400pt}}
\put(458.0,428.0){\rule[-0.200pt]{0.400pt}{4.577pt}}
\put(448.0,428.0){\rule[-0.200pt]{4.818pt}{0.400pt}}
\put(448.0,447.0){\rule[-0.200pt]{4.818pt}{0.400pt}}
\put(370.0,422.0){\rule[-0.200pt]{0.400pt}{4.336pt}}
\put(360.0,422.0){\rule[-0.200pt]{4.818pt}{0.400pt}}
\put(360.0,440.0){\rule[-0.200pt]{4.818pt}{0.400pt}}
\put(339.0,418.0){\rule[-0.200pt]{0.400pt}{6.263pt}}
\put(329.0,418.0){\rule[-0.200pt]{4.818pt}{0.400pt}}
\put(925,474){\circle{18}}
\put(578,434){\circle{18}}
\put(455,440){\circle{18}}
\put(458,438){\circle{18}}
\put(370,431){\circle{18}}
\put(339,431){\circle{18}}
\put(329.0,444.0){\rule[-0.200pt]{4.818pt}{0.400pt}}
\put(300,346){\usebox{\plotpoint}}
\put(300.00,346.00){\usebox{\plotpoint}}
\put(320.69,347.00){\usebox{\plotpoint}}
\put(341.38,348.00){\usebox{\plotpoint}}
\put(362.07,349.00){\usebox{\plotpoint}}
\put(382.76,350.00){\usebox{\plotpoint}}
\put(403.45,351.00){\usebox{\plotpoint}}
\put(424.13,352.00){\usebox{\plotpoint}}
\put(444.82,353.00){\usebox{\plotpoint}}
\put(465.50,354.00){\usebox{\plotpoint}}
\put(486.20,355.00){\usebox{\plotpoint}}
\put(506.89,356.00){\usebox{\plotpoint}}
\put(527.57,357.00){\usebox{\plotpoint}}
\put(548.28,357.78){\usebox{\plotpoint}}
\put(568.99,358.50){\usebox{\plotpoint}}
\put(589.70,359.24){\usebox{\plotpoint}}
\put(610.40,360.00){\usebox{\plotpoint}}
\put(631.08,361.00){\usebox{\plotpoint}}
\put(651.78,362.00){\usebox{\plotpoint}}
\put(672.47,363.00){\usebox{\plotpoint}}
\put(693.15,364.00){\usebox{\plotpoint}}
\put(713.84,365.00){\usebox{\plotpoint}}
\put(734.52,366.00){\usebox{\plotpoint}}
\put(755.22,367.00){\usebox{\plotpoint}}
\put(775.91,368.00){\usebox{\plotpoint}}
\put(796.59,369.00){\usebox{\plotpoint}}
\put(817.28,370.00){\usebox{\plotpoint}}
\put(837.97,371.00){\usebox{\plotpoint}}
\put(858.68,371.71){\usebox{\plotpoint}}
\put(879.39,372.48){\usebox{\plotpoint}}
\put(900.09,373.26){\usebox{\plotpoint}}
\put(920.80,374.00){\usebox{\plotpoint}}
\put(941.49,375.00){\usebox{\plotpoint}}
\put(962.18,376.00){\usebox{\plotpoint}}
\put(982.87,377.00){\usebox{\plotpoint}}
\put(1003.56,378.00){\usebox{\plotpoint}}
\put(1024.25,379.00){\usebox{\plotpoint}}
\put(1044.94,380.00){\usebox{\plotpoint}}
\put(1050,380){\usebox{\plotpoint}}
\put(300,435){\usebox{\plotpoint}}
\put(300.00,435.00){\usebox{\plotpoint}}
\put(320.76,435.00){\usebox{\plotpoint}}
\put(341.51,435.00){\usebox{\plotpoint}}
\put(362.27,435.00){\usebox{\plotpoint}}
\put(383.02,435.00){\usebox{\plotpoint}}
\put(403.78,435.00){\usebox{\plotpoint}}
\put(424.53,435.00){\usebox{\plotpoint}}
\put(445.29,435.00){\usebox{\plotpoint}}
\put(466.04,435.00){\usebox{\plotpoint}}
\put(486.80,435.00){\usebox{\plotpoint}}
\put(507.55,435.00){\usebox{\plotpoint}}
\put(528.31,435.00){\usebox{\plotpoint}}
\put(549.07,435.00){\usebox{\plotpoint}}
\put(569.75,436.00){\usebox{\plotpoint}}
\put(590.51,436.00){\usebox{\plotpoint}}
\put(611.26,436.00){\usebox{\plotpoint}}
\put(632.02,436.00){\usebox{\plotpoint}}
\put(652.77,436.00){\usebox{\plotpoint}}
\put(673.53,436.00){\usebox{\plotpoint}}
\put(694.28,436.00){\usebox{\plotpoint}}
\put(715.04,436.00){\usebox{\plotpoint}}
\put(735.79,436.00){\usebox{\plotpoint}}
\put(756.55,436.00){\usebox{\plotpoint}}
\put(777.31,436.00){\usebox{\plotpoint}}
\put(798.06,436.00){\usebox{\plotpoint}}
\put(818.82,436.00){\usebox{\plotpoint}}
\put(839.57,436.00){\usebox{\plotpoint}}
\put(860.33,436.00){\usebox{\plotpoint}}
\put(881.08,436.00){\usebox{\plotpoint}}
\put(901.84,436.00){\usebox{\plotpoint}}
\put(922.59,436.00){\usebox{\plotpoint}}
\put(943.35,436.00){\usebox{\plotpoint}}
\put(964.04,437.00){\usebox{\plotpoint}}
\put(984.80,437.00){\usebox{\plotpoint}}
\put(1005.55,437.00){\usebox{\plotpoint}}
\put(1026.31,437.00){\usebox{\plotpoint}}
\put(1047.06,437.00){\usebox{\plotpoint}}
\put(1050,437){\usebox{\plotpoint}}
\end{picture}
\end	{center}
\vskip -0.5cm
\caption{The lightest $0^{++}$, $\bullet$, and $2^{++}$, $\circ$,
glueball masses expressed in units of the  string 
tension, in the continuum limit, plotted against $1/N^2$. 
Dotted lines are extrapolations to $N=\infty$.}
\label{fig_gkNwa}
\end 	{figure}
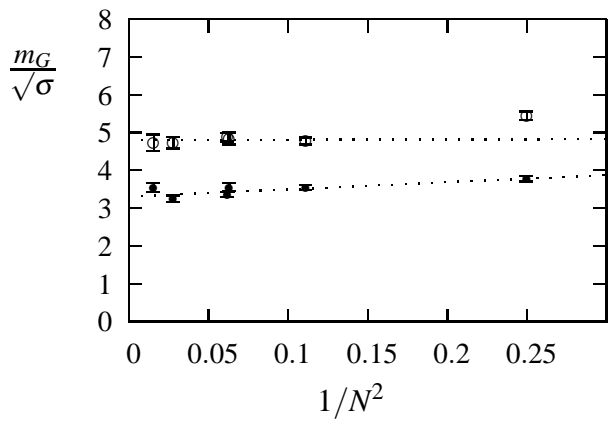

The fact that for many quantities SU(3) $\simeq$ SU($\infty$)
is demonstrated, for example, by calculations of the lightest 
glueball masses
\cite{blmtuw04}, 
as in Fig.~\ref{fig_gkNwa}.

\subsection{Linear confinement in SU(6) and L\"uscher correction}

\begin	{figure}[ht]
\begin	{center}
\leavevmode
\input	{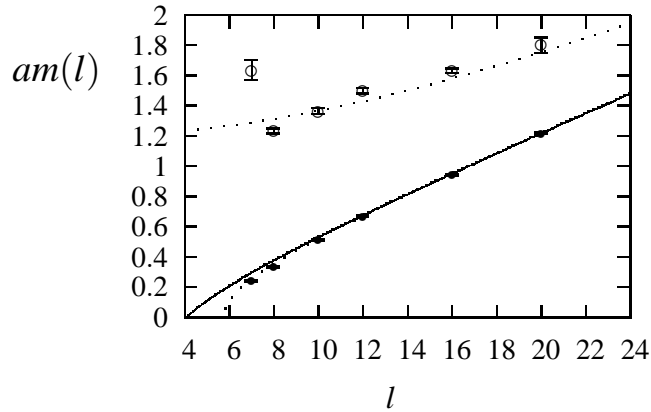}
\end	{center}
\vskip -0.5cm
\caption{The masses of the lightest, $\bullet$, and first
excited, $\circ$, $k=1$ flux loops that wind
around a spatial torus of length $l$ in the SU(6) 
calculation at $\beta=25.05$. The dotted
lines are the predictions of the Nambu-Goto string action.}
\label{fig_K1}
\end 	{figure}

Linear confinement at large $N$ is demonstrated
\cite{hmmtstring}
for SU(6) in Fig.~\ref{fig_K1}. In  Fig.~\ref{fig_gkNwa}
we see that the string tension remains finite in the
$N=\infty$ limit in units of the mass gap. So the
$N=\infty$ theory is indeed linearly confining.
 
\begin	{figure}[h!]
\begin	{center}
\leavevmode
\setlength{\unitlength}{0.240900pt}
\ifx\plotpoint\undefined\newsavebox{\plotpoint}\fi
\sbox{\plotpoint}{\rule[-0.200pt]{0.400pt}{0.400pt}}%
\begin{picture}(1125,675)(0,0)
\font\gnuplot=cmr10 at 12pt
\gnuplot
\sbox{\plotpoint}{\rule[-0.200pt]{0.400pt}{0.400pt}}%
\put(350.0,150.0){\rule[-0.200pt]{4.818pt}{0.400pt}}
\put(325,150){\makebox(0,0)[r]{\ \ {$0$}}}
\put(1030.0,150.0){\rule[-0.200pt]{4.818pt}{0.400pt}}
\put(350.0,218.0){\rule[-0.200pt]{4.818pt}{0.400pt}}
\put(325,218){\makebox(0,0)[r]{\ \ {$0.5$}}}
\put(1030.0,218.0){\rule[-0.200pt]{4.818pt}{0.400pt}}
\put(350.0,286.0){\rule[-0.200pt]{4.818pt}{0.400pt}}
\put(325,286){\makebox(0,0)[r]{\ \ {$1$}}}
\put(1030.0,286.0){\rule[-0.200pt]{4.818pt}{0.400pt}}
\put(350.0,354.0){\rule[-0.200pt]{4.818pt}{0.400pt}}
\put(325,354){\makebox(0,0)[r]{\ \ {$1.5$}}}
\put(1030.0,354.0){\rule[-0.200pt]{4.818pt}{0.400pt}}
\put(350.0,421.0){\rule[-0.200pt]{4.818pt}{0.400pt}}
\put(325,421){\makebox(0,0)[r]{\ \ {$2$}}}
\put(1030.0,421.0){\rule[-0.200pt]{4.818pt}{0.400pt}}
\put(350.0,489.0){\rule[-0.200pt]{4.818pt}{0.400pt}}
\put(325,489){\makebox(0,0)[r]{\ \ {$2.5$}}}
\put(1030.0,489.0){\rule[-0.200pt]{4.818pt}{0.400pt}}
\put(350.0,557.0){\rule[-0.200pt]{4.818pt}{0.400pt}}
\put(325,557){\makebox(0,0)[r]{\ \ {$3$}}}
\put(1030.0,557.0){\rule[-0.200pt]{4.818pt}{0.400pt}}
\put(350.0,625.0){\rule[-0.200pt]{4.818pt}{0.400pt}}
\put(325,625){\makebox(0,0)[r]{\ \ {$3.5$}}}
\put(1030.0,625.0){\rule[-0.200pt]{4.818pt}{0.400pt}}
\put(350.0,150.0){\rule[-0.200pt]{0.400pt}{4.818pt}}
\put(350,100){\makebox(0,0){\ {$4$}}}
\put(350.0,605.0){\rule[-0.200pt]{0.400pt}{4.818pt}}
\put(420.0,150.0){\rule[-0.200pt]{0.400pt}{4.818pt}}
\put(420,100){\makebox(0,0){\ {$6$}}}
\put(420.0,605.0){\rule[-0.200pt]{0.400pt}{4.818pt}}
\put(490.0,150.0){\rule[-0.200pt]{0.400pt}{4.818pt}}
\put(490,100){\makebox(0,0){\ {$8$}}}
\put(490.0,605.0){\rule[-0.200pt]{0.400pt}{4.818pt}}
\put(560.0,150.0){\rule[-0.200pt]{0.400pt}{4.818pt}}
\put(560,100){\makebox(0,0){\ {$10$}}}
\put(560.0,605.0){\rule[-0.200pt]{0.400pt}{4.818pt}}
\put(630.0,150.0){\rule[-0.200pt]{0.400pt}{4.818pt}}
\put(630,100){\makebox(0,0){\ {$12$}}}
\put(630.0,605.0){\rule[-0.200pt]{0.400pt}{4.818pt}}
\put(700.0,150.0){\rule[-0.200pt]{0.400pt}{4.818pt}}
\put(700,100){\makebox(0,0){\ {$14$}}}
\put(700.0,605.0){\rule[-0.200pt]{0.400pt}{4.818pt}}
\put(770.0,150.0){\rule[-0.200pt]{0.400pt}{4.818pt}}
\put(770,100){\makebox(0,0){\ {$16$}}}
\put(770.0,605.0){\rule[-0.200pt]{0.400pt}{4.818pt}}
\put(840.0,150.0){\rule[-0.200pt]{0.400pt}{4.818pt}}
\put(840,100){\makebox(0,0){\ {$18$}}}
\put(840.0,605.0){\rule[-0.200pt]{0.400pt}{4.818pt}}
\put(910.0,150.0){\rule[-0.200pt]{0.400pt}{4.818pt}}
\put(910,100){\makebox(0,0){\ {$20$}}}
\put(910.0,605.0){\rule[-0.200pt]{0.400pt}{4.818pt}}
\put(980.0,150.0){\rule[-0.200pt]{0.400pt}{4.818pt}}
\put(980,100){\makebox(0,0){\ {$22$}}}
\put(980.0,605.0){\rule[-0.200pt]{0.400pt}{4.818pt}}
\put(1050.0,150.0){\rule[-0.200pt]{0.400pt}{4.818pt}}
\put(1050,100){\makebox(0,0){\ {$24$}}}
\put(1050.0,605.0){\rule[-0.200pt]{0.400pt}{4.818pt}}
\put(350.0,150.0){\rule[-0.200pt]{168.630pt}{0.400pt}}
\put(1050.0,150.0){\rule[-0.200pt]{0.400pt}{114.427pt}}
\put(350.0,625.0){\rule[-0.200pt]{168.630pt}{0.400pt}}
\put(150,462){\makebox(0,0){\Large{$c_{eff}$}}}
\put(675,25){\makebox(0,0){\large{$l$}}}
\put(350.0,150.0){\rule[-0.200pt]{0.400pt}{114.427pt}}
\put(469.0,318.0){\rule[-0.200pt]{0.400pt}{13.731pt}}
\put(459.0,318.0){\rule[-0.200pt]{4.818pt}{0.400pt}}
\put(459.0,375.0){\rule[-0.200pt]{4.818pt}{0.400pt}}
\put(522.0,353.0){\rule[-0.200pt]{0.400pt}{8.913pt}}
\put(512.0,353.0){\rule[-0.200pt]{4.818pt}{0.400pt}}
\put(512.0,390.0){\rule[-0.200pt]{4.818pt}{0.400pt}}
\put(592.0,308.0){\rule[-0.200pt]{0.400pt}{12.045pt}}
\put(582.0,308.0){\rule[-0.200pt]{4.818pt}{0.400pt}}
\put(582.0,358.0){\rule[-0.200pt]{4.818pt}{0.400pt}}
\put(697.0,265.0){\rule[-0.200pt]{0.400pt}{9.636pt}}
\put(687.0,265.0){\rule[-0.200pt]{4.818pt}{0.400pt}}
\put(687.0,305.0){\rule[-0.200pt]{4.818pt}{0.400pt}}
\put(837.0,289.0){\rule[-0.200pt]{0.400pt}{19.754pt}}
\put(827.0,289.0){\rule[-0.200pt]{4.818pt}{0.400pt}}
\put(827.0,371.0){\rule[-0.200pt]{4.818pt}{0.400pt}}
\put(452.0,346.0){\rule[-0.200pt]{8.431pt}{0.400pt}}
\put(452.0,336.0){\rule[-0.200pt]{0.400pt}{4.818pt}}
\put(487.0,336.0){\rule[-0.200pt]{0.400pt}{4.818pt}}
\put(487.0,372.0){\rule[-0.200pt]{16.863pt}{0.400pt}}
\put(487.0,362.0){\rule[-0.200pt]{0.400pt}{4.818pt}}
\put(557.0,362.0){\rule[-0.200pt]{0.400pt}{4.818pt}}
\put(557.0,333.0){\rule[-0.200pt]{16.863pt}{0.400pt}}
\put(557.0,323.0){\rule[-0.200pt]{0.400pt}{4.818pt}}
\put(627.0,323.0){\rule[-0.200pt]{0.400pt}{4.818pt}}
\put(627.0,285.0){\rule[-0.200pt]{33.726pt}{0.400pt}}
\put(627.0,275.0){\rule[-0.200pt]{0.400pt}{4.818pt}}
\put(767.0,275.0){\rule[-0.200pt]{0.400pt}{4.818pt}}
\put(767.0,330.0){\rule[-0.200pt]{33.726pt}{0.400pt}}
\put(767.0,320.0){\rule[-0.200pt]{0.400pt}{4.818pt}}
\put(469,346){\circle*{24}}
\put(522,372){\circle*{24}}
\put(592,333){\circle*{24}}
\put(697,285){\circle*{24}}
\put(837,330){\circle*{24}}
\put(907.0,320.0){\rule[-0.200pt]{0.400pt}{4.818pt}}
\put(350,286){\usebox{\plotpoint}}
\put(350.00,286.00){\usebox{\plotpoint}}
\put(370.76,286.00){\usebox{\plotpoint}}
\put(391.51,286.00){\usebox{\plotpoint}}
\put(412.27,286.00){\usebox{\plotpoint}}
\put(433.02,286.00){\usebox{\plotpoint}}
\put(453.78,286.00){\usebox{\plotpoint}}
\put(474.53,286.00){\usebox{\plotpoint}}
\put(495.29,286.00){\usebox{\plotpoint}}
\put(516.04,286.00){\usebox{\plotpoint}}
\put(536.80,286.00){\usebox{\plotpoint}}
\put(557.55,286.00){\usebox{\plotpoint}}
\put(578.31,286.00){\usebox{\plotpoint}}
\put(599.07,286.00){\usebox{\plotpoint}}
\put(619.82,286.00){\usebox{\plotpoint}}
\put(640.58,286.00){\usebox{\plotpoint}}
\put(661.33,286.00){\usebox{\plotpoint}}
\put(682.09,286.00){\usebox{\plotpoint}}
\put(702.84,286.00){\usebox{\plotpoint}}
\put(723.60,286.00){\usebox{\plotpoint}}
\put(744.35,286.00){\usebox{\plotpoint}}
\put(765.11,286.00){\usebox{\plotpoint}}
\put(785.87,286.00){\usebox{\plotpoint}}
\put(806.62,286.00){\usebox{\plotpoint}}
\put(827.38,286.00){\usebox{\plotpoint}}
\put(848.13,286.00){\usebox{\plotpoint}}
\put(868.89,286.00){\usebox{\plotpoint}}
\put(889.64,286.00){\usebox{\plotpoint}}
\put(910.40,286.00){\usebox{\plotpoint}}
\put(931.15,286.00){\usebox{\plotpoint}}
\put(951.91,286.00){\usebox{\plotpoint}}
\put(972.66,286.00){\usebox{\plotpoint}}
\put(993.42,286.00){\usebox{\plotpoint}}
\put(1014.18,286.00){\usebox{\plotpoint}}
\put(1034.93,286.00){\usebox{\plotpoint}}
\put(1050,286){\usebox{\plotpoint}}
\end{picture}
\end	{center}
\vskip -0.5cm
\caption{$am(l) = a^2\sigma l - c_{eff} \frac{\pi}{3}\frac{1}{l}$}
\label{fig_ceff}
\end 	{figure}
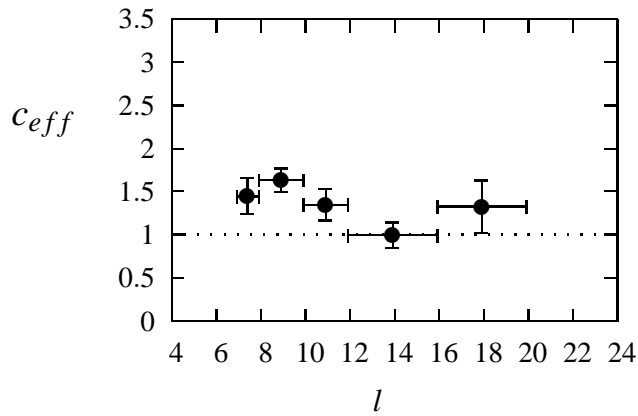

A local fit to the leading string correction, as in 
Fig.~\ref{fig_ceff}, provides good evidence 
that the long-distance behaviour is that of a simple
bosonic string
\cite{hmmtstring}.

\subsection{`t Hooft coupling, $\lambda\equiv g^2N$, fixed
for $N\to\infty$}

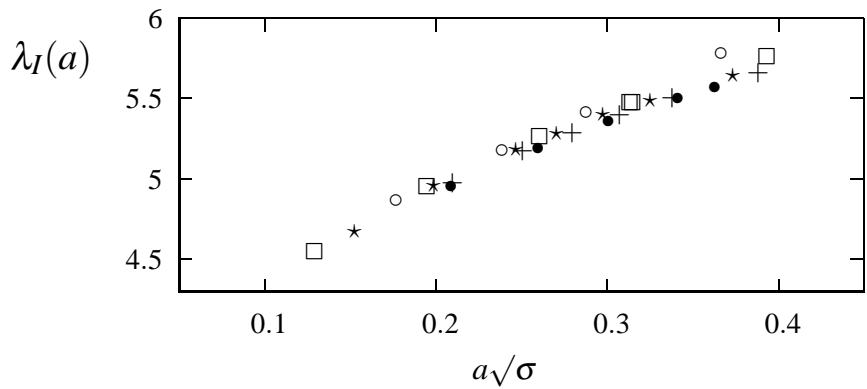
\begin	{figure}[h!]
\begin	{center}
\leavevmode
\setlength{\unitlength}{0.240900pt}
\ifx\plotpoint\undefined\newsavebox{\plotpoint}\fi
\sbox{\plotpoint}{\rule[-0.200pt]{0.400pt}{0.400pt}}%
\begin{picture}(1500,630)(0,0)
\font\gnuplot=cmr10 at 12pt
\gnuplot
\sbox{\plotpoint}{\rule[-0.200pt]{0.400pt}{0.400pt}}%
\put(350.0,201.0){\rule[-0.200pt]{4.818pt}{0.400pt}}
\put(325,201){\makebox(0,0)[r]{\ \ {$4.5$}}}
\put(1405.0,201.0){\rule[-0.200pt]{4.818pt}{0.400pt}}
\put(350.0,327.0){\rule[-0.200pt]{4.818pt}{0.400pt}}
\put(325,327){\makebox(0,0)[r]{\ \ {$5$}}}
\put(1405.0,327.0){\rule[-0.200pt]{4.818pt}{0.400pt}}
\put(350.0,454.0){\rule[-0.200pt]{4.818pt}{0.400pt}}
\put(325,454){\makebox(0,0)[r]{\ \ {$5.5$}}}
\put(1405.0,454.0){\rule[-0.200pt]{4.818pt}{0.400pt}}
\put(350.0,580.0){\rule[-0.200pt]{4.818pt}{0.400pt}}
\put(325,580){\makebox(0,0)[r]{\ \ {$6$}}}
\put(1405.0,580.0){\rule[-0.200pt]{4.818pt}{0.400pt}}
\put(484.0,150.0){\rule[-0.200pt]{0.400pt}{4.818pt}}
\put(484,100){\makebox(0,0){\ {$0.1$}}}
\put(484.0,560.0){\rule[-0.200pt]{0.400pt}{4.818pt}}
\put(753.0,150.0){\rule[-0.200pt]{0.400pt}{4.818pt}}
\put(753,100){\makebox(0,0){\ {$0.2$}}}
\put(753.0,560.0){\rule[-0.200pt]{0.400pt}{4.818pt}}
\put(1022.0,150.0){\rule[-0.200pt]{0.400pt}{4.818pt}}
\put(1022,100){\makebox(0,0){\ {$0.3$}}}
\put(1022.0,560.0){\rule[-0.200pt]{0.400pt}{4.818pt}}
\put(1291.0,150.0){\rule[-0.200pt]{0.400pt}{4.818pt}}
\put(1291,100){\makebox(0,0){\ {$0.4$}}}
\put(1291.0,560.0){\rule[-0.200pt]{0.400pt}{4.818pt}}
\put(350.0,150.0){\rule[-0.200pt]{258.967pt}{0.400pt}}
\put(1425.0,150.0){\rule[-0.200pt]{0.400pt}{103.587pt}}
\put(350.0,580.0){\rule[-0.200pt]{258.967pt}{0.400pt}}
\put(150,515){\makebox(0,0){\Large{$\lambda_I(a)$}}}
\put(862,25){\makebox(0,0){\large{$a\surd\sigma$}}}
\put(350.0,150.0){\rule[-0.200pt]{0.400pt}{103.587pt}}
\put(1191,472){\circle*{18}}
\put(1133,454){\circle*{18}}
\put(1024,418){\circle*{18}}
\put(914,375){\circle*{18}}
\put(777,315){\circle*{18}}
\put(1260,494){\makebox(0,0){$+$}}
\put(1125,455){\makebox(0,0){$+$}}
\put(1042,427){\makebox(0,0){$+$}}
\put(968,400){\makebox(0,0){$+$}}
\put(890,371){\makebox(0,0){$+$}}
\put(780,321){\makebox(0,0){$+$}}
\put(1220,490){\makebox(0,0){$\star$}}
\put(1090,451){\makebox(0,0){$\star$}}
\put(1016,428){\makebox(0,0){$\star$}}
\put(943,398){\makebox(0,0){$\star$}}
\put(879,374){\makebox(0,0){$\star$}}
\put(750,316){\makebox(0,0){$\star$}}
\put(626,245){\makebox(0,0){$\star$}}
\put(1273,518){\raisebox{-.8pt}{\makebox(0,0){$\Box$}}}
\put(1062,446){\raisebox{-.8pt}{\makebox(0,0){$\Box$}}}
\put(1058,445){\raisebox{-.8pt}{\makebox(0,0){$\Box$}}}
\put(916,392){\raisebox{-.8pt}{\makebox(0,0){$\Box$}}}
\put(739,313){\raisebox{-.8pt}{\makebox(0,0){$\Box$}}}
\put(563,211){\raisebox{-.8pt}{\makebox(0,0){$\Box$}}}
\put(1201,525){\circle{18}}
\put(989,432){\circle{18}}
\put(857,373){\circle{18}}
\put(691,294){\circle{18}}
\end{picture}
\end	{center}
\vskip -0.5cm
\caption{The value of the `t Hooft coupling on the scale $a$,
as obtained from mean-field improved $\beta$,
for $N=2(\circ),3(\Box),4(\star),6(+),8(\bullet)$, plotted against the
values of $a$ expressed in physical units.}
\label{fig_ggNI}
\end 	{figure}

We also see 
\cite{blmtuwT05}
in Fig.~\ref{fig_ggNI} that for a smooth large-$N$ limit we need 
to keep  $\lambda(a)  = g^2(a)N$ fixed (at fixed $a\surd\sigma$) 
as expected from diagrams.

\subsection{Pomeron is leading glueball Regge trajectory}

\begin	{figure}[h!]
\begin	{center}
\leavevmode
\input	{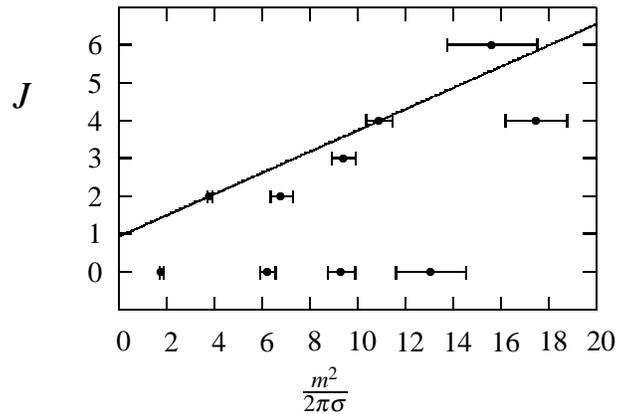}
\end	{center}
\vskip -0.5cm
\caption{Chew-Frautschi plot of $PC=++$ states in the continuum 
SU(3) gauge theory. The leading Regge trajectory is shown.}
\label{fig_pom}
\end 	{figure}

Using novel techniques to calculate masses of high spin glueballs,
one can obtain some solid evidence, as in Fig.~\ref{fig_pom}, for the 
fact that the Pomeron is the leading glueball Regge trajectory 
\cite{hmmtPom}.
This is for SU(3), and is a step towards  $N=\infty$ where mixing and
decay ambiguities disappear.

\subsection{Topology}

\begin  {figure}[h!]
\begin  {center}
\leavevmode
\setlength{\unitlength}{0.240900pt}
\ifx\plotpoint\undefined\newsavebox{\plotpoint}\fi
\sbox{\plotpoint}{\rule[-0.200pt]{0.400pt}{0.400pt}}%
\begin{picture}(1275,675)(0,0)
\font\gnuplot=cmr10 at 12pt
\gnuplot
\sbox{\plotpoint}{\rule[-0.200pt]{0.400pt}{0.400pt}}%
\put(375.0,150.0){\rule[-0.200pt]{4.818pt}{0.400pt}}
\put(350,150){\makebox(0,0)[r]{\ \ {$0$}}}
\put(1180.0,150.0){\rule[-0.200pt]{4.818pt}{0.400pt}}
\put(375.0,229.0){\rule[-0.200pt]{4.818pt}{0.400pt}}
\put(350,229){\makebox(0,0)[r]{\ \ {$500$}}}
\put(1180.0,229.0){\rule[-0.200pt]{4.818pt}{0.400pt}}
\put(375.0,308.0){\rule[-0.200pt]{4.818pt}{0.400pt}}
\put(350,308){\makebox(0,0)[r]{\ \ {$1000$}}}
\put(1180.0,308.0){\rule[-0.200pt]{4.818pt}{0.400pt}}
\put(375.0,388.0){\rule[-0.200pt]{4.818pt}{0.400pt}}
\put(350,388){\makebox(0,0)[r]{\ \ {$1500$}}}
\put(1180.0,388.0){\rule[-0.200pt]{4.818pt}{0.400pt}}
\put(375.0,467.0){\rule[-0.200pt]{4.818pt}{0.400pt}}
\put(350,467){\makebox(0,0)[r]{\ \ {$2000$}}}
\put(1180.0,467.0){\rule[-0.200pt]{4.818pt}{0.400pt}}
\put(375.0,546.0){\rule[-0.200pt]{4.818pt}{0.400pt}}
\put(350,546){\makebox(0,0)[r]{\ \ {$2500$}}}
\put(1180.0,546.0){\rule[-0.200pt]{4.818pt}{0.400pt}}
\put(375.0,625.0){\rule[-0.200pt]{4.818pt}{0.400pt}}
\put(350,625){\makebox(0,0)[r]{\ \ {$3000$}}}
\put(1180.0,625.0){\rule[-0.200pt]{4.818pt}{0.400pt}}
\put(375.0,150.0){\rule[-0.200pt]{0.400pt}{4.818pt}}
\put(375,100){\makebox(0,0){\ {$0$}}}
\put(375.0,605.0){\rule[-0.200pt]{0.400pt}{4.818pt}}
\put(444.0,150.0){\rule[-0.200pt]{0.400pt}{4.818pt}}
\put(444,100){\makebox(0,0){\ {$1$}}}
\put(444.0,605.0){\rule[-0.200pt]{0.400pt}{4.818pt}}
\put(513.0,150.0){\rule[-0.200pt]{0.400pt}{4.818pt}}
\put(513,100){\makebox(0,0){\ {$2$}}}
\put(513.0,605.0){\rule[-0.200pt]{0.400pt}{4.818pt}}
\put(581.0,150.0){\rule[-0.200pt]{0.400pt}{4.818pt}}
\put(581,100){\makebox(0,0){\ {$3$}}}
\put(581.0,605.0){\rule[-0.200pt]{0.400pt}{4.818pt}}
\put(650.0,150.0){\rule[-0.200pt]{0.400pt}{4.818pt}}
\put(650,100){\makebox(0,0){\ {$4$}}}
\put(650.0,605.0){\rule[-0.200pt]{0.400pt}{4.818pt}}
\put(719.0,150.0){\rule[-0.200pt]{0.400pt}{4.818pt}}
\put(719,100){\makebox(0,0){\ {$5$}}}
\put(719.0,605.0){\rule[-0.200pt]{0.400pt}{4.818pt}}
\put(788.0,150.0){\rule[-0.200pt]{0.400pt}{4.818pt}}
\put(788,100){\makebox(0,0){\ {$6$}}}
\put(788.0,605.0){\rule[-0.200pt]{0.400pt}{4.818pt}}
\put(856.0,150.0){\rule[-0.200pt]{0.400pt}{4.818pt}}
\put(856,100){\makebox(0,0){\ {$7$}}}
\put(856.0,605.0){\rule[-0.200pt]{0.400pt}{4.818pt}}
\put(925.0,150.0){\rule[-0.200pt]{0.400pt}{4.818pt}}
\put(925,100){\makebox(0,0){\ {$8$}}}
\put(925.0,605.0){\rule[-0.200pt]{0.400pt}{4.818pt}}
\put(994.0,150.0){\rule[-0.200pt]{0.400pt}{4.818pt}}
\put(994,100){\makebox(0,0){\ {$9$}}}
\put(994.0,605.0){\rule[-0.200pt]{0.400pt}{4.818pt}}
\put(1063.0,150.0){\rule[-0.200pt]{0.400pt}{4.818pt}}
\put(1063,100){\makebox(0,0){\ {$10$}}}
\put(1063.0,605.0){\rule[-0.200pt]{0.400pt}{4.818pt}}
\put(1131.0,150.0){\rule[-0.200pt]{0.400pt}{4.818pt}}
\put(1131,100){\makebox(0,0){\ {$11$}}}
\put(1131.0,605.0){\rule[-0.200pt]{0.400pt}{4.818pt}}
\put(1200.0,150.0){\rule[-0.200pt]{0.400pt}{4.818pt}}
\put(1200,100){\makebox(0,0){\ {$12$}}}
\put(1200.0,605.0){\rule[-0.200pt]{0.400pt}{4.818pt}}
\put(375.0,150.0){\rule[-0.200pt]{198.742pt}{0.400pt}}
\put(1200.0,150.0){\rule[-0.200pt]{0.400pt}{114.427pt}}
\put(375.0,625.0){\rule[-0.200pt]{198.742pt}{0.400pt}}
\put(150,537){\makebox(0,0){\Large{$D(\rho)$}}}
\put(762,25){\makebox(0,0){\large{$\rho$}}}
\put(375.0,150.0){\rule[-0.200pt]{0.400pt}{114.427pt}}
\put(461,150){\circle*{12}}
\put(495,178){\circle*{12}}
\put(530,237){\circle*{12}}
\put(564,263){\circle*{12}}
\put(598,276){\circle*{12}}
\put(633,271){\circle*{12}}
\put(667,260){\circle*{12}}
\put(702,235){\circle*{12}}
\put(736,210){\circle*{12}}
\put(770,193){\circle*{12}}
\put(805,183){\circle*{12}}
\put(839,171){\circle*{12}}
\put(873,166){\circle*{12}}
\put(908,162){\circle*{12}}
\put(942,157){\circle*{12}}
\put(977,153){\circle*{12}}
\put(1011,152){\circle*{12}}
\put(1045,151){\circle*{12}}
\put(495,154){\makebox(0,0){$\times$}}
\put(530,177){\makebox(0,0){$\times$}}
\put(564,219){\makebox(0,0){$\times$}}
\put(598,292){\makebox(0,0){$\times$}}
\put(633,361){\makebox(0,0){$\times$}}
\put(667,359){\makebox(0,0){$\times$}}
\put(702,306){\makebox(0,0){$\times$}}
\put(736,257){\makebox(0,0){$\times$}}
\put(770,203){\makebox(0,0){$\times$}}
\put(805,183){\makebox(0,0){$\times$}}
\put(839,165){\makebox(0,0){$\times$}}
\put(873,160){\makebox(0,0){$\times$}}
\put(908,154){\makebox(0,0){$\times$}}
\put(942,152){\makebox(0,0){$\times$}}
\put(977,150){\makebox(0,0){$\times$}}
\put(564,154){\circle{18}}
\put(598,228){\circle{18}}
\put(633,443){\circle{18}}
\put(667,570){\circle{18}}
\put(702,410){\circle{18}}
\put(736,284){\circle{18}}
\put(770,212){\circle{18}}
\put(805,181){\circle{18}}
\put(839,163){\circle{18}}
\put(873,155){\circle{18}}
\put(908,152){\circle{18}}
\put(942,151){\circle{18}}
\put(1011,150){\circle{18}}
\put(1045,150){\circle{18}}
\end{picture}
\end    {center}
\vskip -0.5cm
\caption{The `instanton' size density, $D(\rho)$, for
$N=3(\bullet),6(\times),12(\circ)$ at $a \simeq 1/4.5T_c$.}
\label{fig_drho8}
\end    {figure}
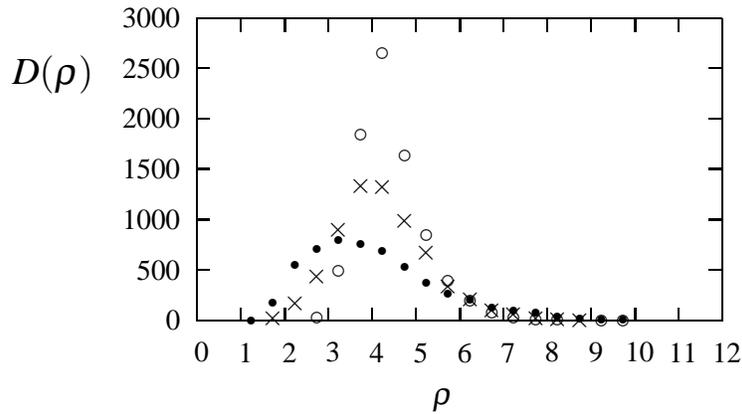

The instanton size distribution seems to head to
$D(\rho)\stackrel{N\to\infty}{\longrightarrow} 
\delta (\rho-\rho_c)$ 
\cite{mtU,blmtuwQ}
as in Fig.~\ref{fig_drho8}.

Also:

$\bullet$ topological susceptibility at $N\to\infty$: 
\cite{pisa1,ox1}

$\bullet$ no topological fluctuations at $T>T_c$ at large $N$: 
\cite{blmtuwQ,pisa2}

$\bullet$ evidence for interlacing $\theta$-vacua: 
\cite{pisa1}

$\bullet$ evidence that topology drives chiral symmetry breaking:
\cite{oxGW}

\subsection{Deconfinement}

\begin	{figure}[ht]
\begin	{center}
\leavevmode
\setlength{\unitlength}{0.240900pt}
\ifx\plotpoint\undefined\newsavebox{\plotpoint}\fi
\sbox{\plotpoint}{\rule[-0.200pt]{0.400pt}{0.400pt}}%
\begin{picture}(1500,675)(0,0)
\font\gnuplot=cmr10 at 12pt
\gnuplot
\sbox{\plotpoint}{\rule[-0.200pt]{0.400pt}{0.400pt}}%
\put(350.0,150.0){\rule[-0.200pt]{4.818pt}{0.400pt}}
\put(325,150){\makebox(0,0)[r]{\ \ {$0$}}}
\put(1405.0,150.0){\rule[-0.200pt]{4.818pt}{0.400pt}}
\put(350.0,198.0){\rule[-0.200pt]{4.818pt}{0.400pt}}
\put(325,198){\makebox(0,0)[r]{\ \ {$0.1$}}}
\put(1405.0,198.0){\rule[-0.200pt]{4.818pt}{0.400pt}}
\put(350.0,245.0){\rule[-0.200pt]{4.818pt}{0.400pt}}
\put(325,245){\makebox(0,0)[r]{\ \ {$0.2$}}}
\put(1405.0,245.0){\rule[-0.200pt]{4.818pt}{0.400pt}}
\put(350.0,293.0){\rule[-0.200pt]{4.818pt}{0.400pt}}
\put(325,293){\makebox(0,0)[r]{\ \ {$0.3$}}}
\put(1405.0,293.0){\rule[-0.200pt]{4.818pt}{0.400pt}}
\put(350.0,340.0){\rule[-0.200pt]{4.818pt}{0.400pt}}
\put(325,340){\makebox(0,0)[r]{\ \ {$0.4$}}}
\put(1405.0,340.0){\rule[-0.200pt]{4.818pt}{0.400pt}}
\put(350.0,388.0){\rule[-0.200pt]{4.818pt}{0.400pt}}
\put(325,388){\makebox(0,0)[r]{\ \ {$0.5$}}}
\put(1405.0,388.0){\rule[-0.200pt]{4.818pt}{0.400pt}}
\put(350.0,435.0){\rule[-0.200pt]{4.818pt}{0.400pt}}
\put(325,435){\makebox(0,0)[r]{\ \ {$0.6$}}}
\put(1405.0,435.0){\rule[-0.200pt]{4.818pt}{0.400pt}}
\put(350.0,483.0){\rule[-0.200pt]{4.818pt}{0.400pt}}
\put(325,483){\makebox(0,0)[r]{\ \ {$0.7$}}}
\put(1405.0,483.0){\rule[-0.200pt]{4.818pt}{0.400pt}}
\put(350.0,530.0){\rule[-0.200pt]{4.818pt}{0.400pt}}
\put(325,530){\makebox(0,0)[r]{\ \ {$0.8$}}}
\put(1405.0,530.0){\rule[-0.200pt]{4.818pt}{0.400pt}}
\put(350.0,578.0){\rule[-0.200pt]{4.818pt}{0.400pt}}
\put(325,578){\makebox(0,0)[r]{\ \ {$0.9$}}}
\put(1405.0,578.0){\rule[-0.200pt]{4.818pt}{0.400pt}}
\put(350.0,625.0){\rule[-0.200pt]{4.818pt}{0.400pt}}
\put(325,625){\makebox(0,0)[r]{\ \ {$1$}}}
\put(1405.0,625.0){\rule[-0.200pt]{4.818pt}{0.400pt}}
\put(350.0,150.0){\rule[-0.200pt]{0.400pt}{4.818pt}}
\put(350,100){\makebox(0,0){\ {$0$}}}
\put(350.0,605.0){\rule[-0.200pt]{0.400pt}{4.818pt}}
\put(529.0,150.0){\rule[-0.200pt]{0.400pt}{4.818pt}}
\put(529,100){\makebox(0,0){\ {$0.05$}}}
\put(529.0,605.0){\rule[-0.200pt]{0.400pt}{4.818pt}}
\put(708.0,150.0){\rule[-0.200pt]{0.400pt}{4.818pt}}
\put(708,100){\makebox(0,0){\ {$0.1$}}}
\put(708.0,605.0){\rule[-0.200pt]{0.400pt}{4.818pt}}
\put(888.0,150.0){\rule[-0.200pt]{0.400pt}{4.818pt}}
\put(888,100){\makebox(0,0){\ {$0.15$}}}
\put(888.0,605.0){\rule[-0.200pt]{0.400pt}{4.818pt}}
\put(1067.0,150.0){\rule[-0.200pt]{0.400pt}{4.818pt}}
\put(1067,100){\makebox(0,0){\ {$0.2$}}}
\put(1067.0,605.0){\rule[-0.200pt]{0.400pt}{4.818pt}}
\put(1246.0,150.0){\rule[-0.200pt]{0.400pt}{4.818pt}}
\put(1246,100){\makebox(0,0){\ {$0.25$}}}
\put(1246.0,605.0){\rule[-0.200pt]{0.400pt}{4.818pt}}
\put(1425.0,150.0){\rule[-0.200pt]{0.400pt}{4.818pt}}
\put(1425,100){\makebox(0,0){\ {$0.3$}}}
\put(1425.0,605.0){\rule[-0.200pt]{0.400pt}{4.818pt}}
\put(350.0,150.0){\rule[-0.200pt]{258.967pt}{0.400pt}}
\put(1425.0,150.0){\rule[-0.200pt]{0.400pt}{114.427pt}}
\put(350.0,625.0){\rule[-0.200pt]{258.967pt}{0.400pt}}
\put(150,537){\makebox(0,0){\Large{${{T_c}\over{\surd\sigma}}$}}}
\put(862,25){\makebox(0,0){\large{$1/N^2$}}}
\put(350.0,150.0){\rule[-0.200pt]{0.400pt}{114.427pt}}
\put(1246.0,485.0){\rule[-0.200pt]{0.400pt}{0.964pt}}
\put(1236.0,485.0){\rule[-0.200pt]{4.818pt}{0.400pt}}
\put(1236.0,489.0){\rule[-0.200pt]{4.818pt}{0.400pt}}
\put(748.0,456.0){\rule[-0.200pt]{0.400pt}{0.482pt}}
\put(738.0,456.0){\rule[-0.200pt]{4.818pt}{0.400pt}}
\put(738.0,458.0){\rule[-0.200pt]{4.818pt}{0.400pt}}
\put(574.0,449.0){\rule[-0.200pt]{0.400pt}{0.964pt}}
\put(564.0,449.0){\rule[-0.200pt]{4.818pt}{0.400pt}}
\put(564.0,453.0){\rule[-0.200pt]{4.818pt}{0.400pt}}
\put(450.0,437.0){\rule[-0.200pt]{0.400pt}{1.204pt}}
\put(440.0,437.0){\rule[-0.200pt]{4.818pt}{0.400pt}}
\put(440.0,442.0){\rule[-0.200pt]{4.818pt}{0.400pt}}
\put(406.0,426.0){\rule[-0.200pt]{0.400pt}{2.650pt}}
\put(396.0,426.0){\rule[-0.200pt]{4.818pt}{0.400pt}}
\put(1246,487){\circle*{12}}
\put(748,457){\circle*{12}}
\put(574,451){\circle*{12}}
\put(450,440){\circle*{12}}
\put(406,432){\circle*{12}}
\put(396.0,437.0){\rule[-0.200pt]{4.818pt}{0.400pt}}
\put(350,434){\usebox{\plotpoint}}
\put(350.00,434.00){\usebox{\plotpoint}}
\put(370.72,434.88){\usebox{\plotpoint}}
\put(391.42,436.00){\usebox{\plotpoint}}
\put(412.10,437.74){\usebox{\plotpoint}}
\put(432.81,438.62){\usebox{\plotpoint}}
\put(453.51,440.00){\usebox{\plotpoint}}
\put(474.21,441.00){\usebox{\plotpoint}}
\put(494.90,442.36){\usebox{\plotpoint}}
\put(515.62,443.24){\usebox{\plotpoint}}
\put(536.30,445.00){\usebox{\plotpoint}}
\put(557.00,446.09){\usebox{\plotpoint}}
\put(577.72,447.00){\usebox{\plotpoint}}
\put(598.39,448.85){\usebox{\plotpoint}}
\put(619.10,449.81){\usebox{\plotpoint}}
\put(639.81,450.71){\usebox{\plotpoint}}
\put(660.51,452.00){\usebox{\plotpoint}}
\put(681.20,453.47){\usebox{\plotpoint}}
\put(701.91,454.45){\usebox{\plotpoint}}
\put(722.59,456.00){\usebox{\plotpoint}}
\put(743.29,457.21){\usebox{\plotpoint}}
\put(764.01,458.10){\usebox{\plotpoint}}
\put(784.67,460.00){\usebox{\plotpoint}}
\put(805.39,460.94){\usebox{\plotpoint}}
\put(826.10,461.83){\usebox{\plotpoint}}
\put(846.80,463.00){\usebox{\plotpoint}}
\put(867.48,464.68){\usebox{\plotpoint}}
\put(888.20,465.56){\usebox{\plotpoint}}
\put(908.89,467.00){\usebox{\plotpoint}}
\put(929.58,468.36){\usebox{\plotpoint}}
\put(950.29,469.30){\usebox{\plotpoint}}
\put(971.01,470.18){\usebox{\plotpoint}}
\put(991.68,472.00){\usebox{\plotpoint}}
\put(1012.38,473.03){\usebox{\plotpoint}}
\put(1033.09,474.00){\usebox{\plotpoint}}
\put(1053.77,475.80){\usebox{\plotpoint}}
\put(1074.48,476.68){\usebox{\plotpoint}}
\put(1095.18,478.00){\usebox{\plotpoint}}
\put(1115.89,479.00){\usebox{\plotpoint}}
\put(1136.58,480.42){\usebox{\plotpoint}}
\put(1157.29,481.33){\usebox{\plotpoint}}
\put(1177.97,483.00){\usebox{\plotpoint}}
\put(1198.67,484.15){\usebox{\plotpoint}}
\put(1219.38,485.03){\usebox{\plotpoint}}
\put(1240.05,487.00){\usebox{\plotpoint}}
\put(1260.76,487.89){\usebox{\plotpoint}}
\put(1281.48,488.77){\usebox{\plotpoint}}
\put(1302.18,490.00){\usebox{\plotpoint}}
\put(1322.85,491.62){\usebox{\plotpoint}}
\put(1343.57,492.51){\usebox{\plotpoint}}
\put(1364.26,494.00){\usebox{\plotpoint}}
\put(1384.95,495.30){\usebox{\plotpoint}}
\put(1405.66,496.24){\usebox{\plotpoint}}
\put(1425,497){\usebox{\plotpoint}}
\end{picture}
\end	{center}
\vskip -0.5cm
\caption{The deconfining temperature in units of the
string tension for various SU($N$) gauge theories.
Large $N$ extrapolation shown.}
\label{fig_tcN}
\end 	{figure}
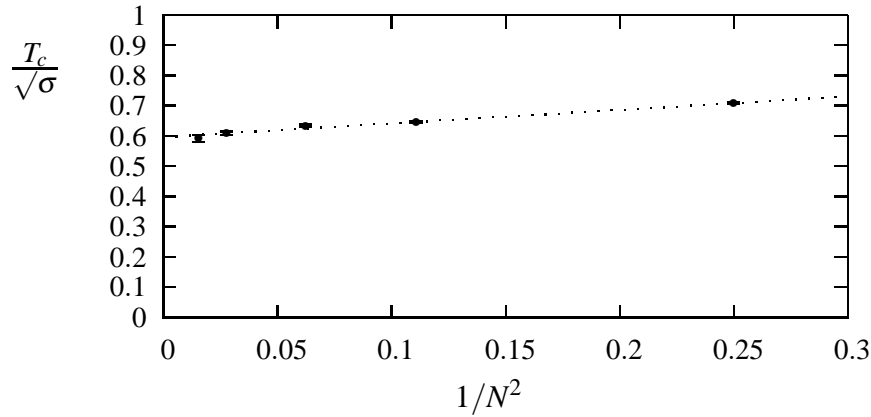

$T_c$ rapidly converges to its large-$N$ limit,
as in Fig.~\ref{fig_tcN}, becoming more
strongly first order as we see from the latent heat
plot in 
\cite{blmtuwT05,blmtuwQ,blmtuwT03}.

\subsection{$k$-strings}

New stable confining strings appear at larger $N$
\cite{blmtuw04,blmt00,blmt01,pisa01b,pisa01a}
and their ratios, as listed 
in Table~\ref{table_ksig} from \cite{blmtuw04},
can be compared to the Casimir Scaling and
MQCD-inspired conjectures.


\begin{table}[h!]
\begin{center}
\begin{tabular}{|c||c|c|c|}\hline
\multicolumn{4}{|c|}{ $\sigma_{k}/\sigma$ } \\ \hline
 (N,k) & Casimir scaling & this paper & `MQCD' \\ \hline
(4,2) & 1.333 & 1.370(20) &  1.414 \\
(4,2) & 1.333 & 1.358(33) &  1.414 \\
(6,2) & 1.600 & 1.675(31) &  1.732 \\
(6,3) & 1.800 & 1.886(61) &  2.000 \\
(8,2) & 1.714 & 1.779(51) &  1.848 \\
(8,3) & 2.143 & 2.38(10)  &  2.414 \\
(8,4) & 2.286 & 2.69(17)  &  2.613 \\ \hline
\end{tabular}
\caption{\label{table_ksig}
Predictions of `Casimir Scaling' and `MQCD' compared against
calculated values of
the ratio of the tension of the lightest $k$-string
to that of the fundamental ($k=1$) string. The second
SU(4) calculation is on anisotropic lattices.}
\end{center}
\end{table}

\subsection{.... and more ....}

Pressure deficit above $T_c$ at large $N$
\cite{bbmtP}.
Hunting the Hagedorn phase transition
\cite{bbmtH}.
Large-$N$ phases
\cite{NN,hv}.
't Hooft string tensions
\cite{fbmtDW,pdfDW}.
D=2+1 deconfinement at all $N$
\cite{jlD3}.
Space-time reduction at large $N$
\cite{hv}.
Chiral symmetry and quark masses at large $N$
\cite{NN2}
...
Mesons and baryons at large $N$
...
${\cal{N}}=1$ SUSY at $N=\infty$ 
....

\end{document}